# Correlation between Complex Spin Textures and the Magnetocaloric and Hall Effects in Eu(Ga$_{1-x}$Al$_x$)$_4$ ($x$ = 0.9, 1)

Kelly J. Neubauer[1,2,3,*], Kevin Allen[1,2,3], Jaime M. Moya[1,4,5], Mason L. Klemm[1,2,3], Feng Ye[6], Zachary Morgan[6], Lisa DeBeer-Schmitt[6], Wei Tian[6], Emilia Morosan[1,2,3], Pengcheng Dai[1,2,3,**]

[1]Department of Physics and Astronomy, Rice University, Houston, TX, 77005, USA
[2]Smalley-Curl Institute, Rice University, Houston, TX, 77005, USA
[3]Rice Center for Quantum Materials, Rice University, Houston, TX, 77005, USA
[4]Applied Physics Graduate Program, Rice University, Houston, Texas 77005, USA
[5]Department of Chemistry, Princeton University, Princeton, NJ 08544, USA
[6]Neutron Scattering Division, Oak Ridge National Laboratory, Oak Ridge, TN, 37831, USA
*email: kneu28@gmail.com; **email: pdai@rice.edu

**Abstract:**
Determining the electronic phase diagram of a quantum material as a function of temperature ($T$) and applied magnetic field ($H$) forms the basis for understanding the microscopic origin of transport properties, such as the anomalous Hall effect (AHE) and topological Hall effect (THE). For many magnetic quantum materials, including EuAl$_4$, a THE arises from a topologically protected magnetic skyrmion lattice with a non-zero scalar spin chirality. We identified a square skyrmion lattice (sSkL) peak in Eu(Ga$_{1-x}$Al$_x$)$_4$ ($x$ = 0.9) identical to the peak previously observed in EuAl$_4$ by performing neutron scattering measurements throughout the phase diagram. Comparing these neutron results with transport measurements, we found that in both compounds the maximal THE does not correspond to the sSkL area. Instead of the maximal THE, the maximal magnetocaloric effect (MCE) boundaries better identify the sSkL lattice phase observed by neutron scattering measurements. The maximal THE therefore arises from interactions of itinerant electrons with frustrated spin fluctuations in a topologically trivial magnetic state.

## I.     INTRODUCTION

The conventional Hall effect is the potential difference ($V$) across a nonmagnetic electric conductor transverse to the current ($I$) and to an external magnetic field ($H$) perpendicular to $I$ [1]. The unconventional Hall effect includes anomalous, topological, and spin components and describes additional effects from magnetism in conducting quantum materials not included in the conventional Hall effect.  While the conventional anomalous Hall effect (AHE) was initially observed in ferromagnetic (FM) materials with spontaneous magnetization ($M$) [1], AHEs have since been observed in antiferromagnetic (AFM) materials due to the interplay of electrons and spins [2-4]. Generally, the AHE is given by $\rho_{yx}^A = \rho_{yx}^{A'} + \rho_{yx}^T$, where the first contribution, $\rho_{yx}^{A'}$, represents the conventional AHE and scales with the net magnetization. The second contribution, $\rho_{yx}^T$, represents the topological Hall effect (THE). In contrast to the conventional AHE, the THE is not proportional to a net magnetization and has been considered the signature of spin textures with nonzero scalar spin chirality (SSC) [$\chi = \boldsymbol{S}_i \cdot (\boldsymbol{S}_j \times \boldsymbol{S}_k) \neq 0$, where $\boldsymbol{S}_i$, $\boldsymbol{S}_j$, $\boldsymbol{S}_k$ are the three nearest spins] in real space [5-8]. Noncoplanar spin structures, such as skyrmions, have nonzero SSC and exhibit nonzero Berry curvature that acts as an effective magnetic field and gives rise to a topological contribution to the Hall resistivity. Since skyrmions can be easily manipulated, they hold tremendous promise for novel spintronics applications [8-11].

Most topological spin textures and corresponding THE have been realized in noncentrosymmetric materials such as MnSi and related systems where the Dzyaloshinskii-Moriya (DM) interaction stabilizes skyrmions [5, 12-17]. Recently, several centrosymmetric materials have been reported to host skyrmion lattices. This includes the centrosymmetric Eu-based materials $EuAl_2Ga_2$ [18] and $EuAl_4$ (Fig. 1a) [14] and Gd-based compounds such as $GdRu_2Si_2$ [19], $Gd_3Ru_4Al_{12}$ [6], and $Gd_2PdSi_3$ [7, 20, 21]. These materials host incommensurate ultra-small skyrmions stabilized with magnetic fields, as illustrated in Fig. 1b [22]. In centrosymmetric systems, the underlying mechanism of the skyrmions and their relation to the magnetotransport effects are not well understood. In particular, it is unclear how these skyrmions contribute to the AHE compared with the DM- or geometric frustration-induced skyrmions in noncentrosymmetric materials [5, 12-17].

In addition to magnetic skyrmion lattice [8], the AHE or THE can also arise from conduction electrons scattering off spin fluctuations in geometrically-frustrated antiferromagnets, as seen in the kagome lattice antiferromagnet $YMn_6Sn_6$ [23-25] and the Gd-based compounds $Gd_3Ru_4Al_{12}$ (triangular lattice) [6] and $Gd_2PdSi_3$ (breathing Kagome) [7, 20, 21]. Furthermore, metamagnetic multiband effects can give an AHE-like signal in the Ising antiferromagnet $ErGa_2$ [26]. Since spin fluctuations in these antiferromagnets and metamagnetic multiband effects are topologically trivial and clearly different from topologically nontrivial magnetic skyrmions [8], it is important to identify the contribution of magnetic skyrmions to the AHE.

The $Eu(Ga_{1-x}Al_x)_4$ system is particularly interesting among centrosymmetric skyrmion materials. $Eu(Ga_{1-x}Al_x)_4$ compounds are tetragonal with space group I4/mmm as shown in Fig. 1(a). For $x = 1$, an incommensurate charge-density-wave (CDW) order appears around 140 K with modulation wave vector $Q_{CDW} = (0, 0, 0.183)$ along the $c$-axis [27]. The $Q_{CDW}$ gradually shifts to a commensurate position of (0, 0, 1/6) with decreasing temperature [27]. Very recently, resonant magnetic X-ray scattering (XRMS) experiments on $EuAl_4$ unveiled a first-order transition associated with a spontaneous chirality flipping of the helical magnetic order at 10.1 K [27, 28]. Intriguingly, the CDW order strongly couples with the magnetic order and exhibits a commensurate-to-incommensurate transition at the chirality flipping temperature [27], much different from the CDW and magnetic order coupling in kagome lattice antiferromagnet FeGe [29]. A similar coupling of the CDW and magnetic order was also reported in the related compound $EuGa_2Al_2$ [18].

$Eu(Ga_{1-x}Al_x)_4$ with $x = 1$ orders antiferromagnetically below $T_N = 15.4$ K, followed, on cooling, by three additional magnetic transitions at 13.3 K, 12.3 K, and 10.1 K [27]. $EuAl_4$ has been shown to exhibit a series of metamagnetic transitions when a magnetic field is applied along the $c$-axis, as shown in Fig. 1(c), although the nature of these transitions has been unclear [30]. Additionally, $EuAl_4$ has been shown to host magnetic field-induced square and rhombic skyrmion lattices and exhibits a large AHE [14]. Interestingly, we observe that the maximal THE and skyrmion phases do not occur simultaneously in $EuAl_4$ as it does in other skyrmion systems [14, 31], thus raising the issue concerning the microscopic origin of the oberseved THE and its relationship with the skyrmion phase. Therefore, the $Eu(Ga_{1-x}Al_x)_4$ system provides a unique example of a skyrmion phase that cannot be easily identified based solely on a maximal THE measured via transport.

Upon replacing Al in EuAl$_4$ by Ga to form EuGa$_4$, the structural properties or magnetic ordering temperature do not change substantially [32, 33]. For partially doped Eu(Ga$_{1-x}$Al$_x$)$_4$, previous work showed antiferromagnetic (AFM) ordering with the Neél temperature $T_N$ ranging from 12 to 20 K for $x = 0 – 1$ [32, 34-36]. Despite that, a small 10% Ga doping strongly influences the phase diagram, as shown in Fig. 1(e-f), which hosts a strong and broad maximal THE [35]. Yet the relationship between the THE and spin textures in the 10% doped sample is unknown. Since small angle neutron scattering (SANS) experiments on EuAl$_4$ have previously identified the temperature and applied magnetic field regimes where a skyrmion lattice occurs [14], similar measurements on Ga-doped compounds Eu(Ga$_{1-x}$Al$_x$)$_4$ can determine the evolution of the skyrmion lattice with $x$. By carefully comparing SANS measurements with the maximal THE, we can determine if the maximal THE in centrosymmetric materials is associated with a field-induced skyrmion lattice. If similar to the parent compound, the maximal THE and skyrmion region do not coexist in the same place of the $H$-$T$ phase diagram, we can consider other probes, such as the magnetocaloric effect (MCE), to more accurately locate the skyrmion state in $H$-$T$ space. Although MCE is not directly related to the topological nature of the skyrmion lattice, it is sensitive to the magnetic phase boundary and changes in magnetic entropy induced by the applied magnetic field.

In this paper, we carried out detailed magnetization, magnetotransport, MCE, SANS, and wide angle neutron diffraction measurements on single crystals of Eu(Ga$_{1-x}$Al$_x$)$_4$ with $x = 0.9$ and 1. Our work focuses on the $x = 0.9$ compound, which has previously not been studied via neutron scattering. We determined the magnetic states from neutron scattering measurements. Commensurate (CM) nuclear Bragg peaks, shown in Fig. 1(g), are observed at all temperatures and magnetic fields, while incommensurate (ICM) peaks, shown in Fig. 1(h), appear below $T_N$ = 16 K. For $H \| c$, in a small region of temperature and $c$-axis magnetic field phase space, an additional peak appears, as shown in Fig. 1(i), and corresponds to the emergence of a square skyrmion lattice (sSkL) [14]. The skyrmion lattice occurs at temperatures close to $T_N$, well separated from the maximal THE observed below ~4 K. Therefore, in Eu(Ga$_{0.1}$Al$_{0.9}$)$_4$, a clear mismatch exists between the skyrmion lattice and maximal THE. This result indicates that the skyrmion lattice in Eu(Ga$_{1-x}$Al$_x$)$_4$ is highly sensitive to $x$ and that the maximal THE, typically used to indicate a possible skyrmion state, is not a strong indicator of skyrmions in this system. Instead, we show that the maximal MCE is a strong indicator of the skyrmion phase in this system, providing an early example of a skyrmion lattice phase identified based on its DC magnetization response [37]. Therefore, the maximal THE exists without a topologically nontrivial magnetic state, suggesting that the topological skyrmion lattice phase does not contribute significantly to the THE in centrosymmetric Eu(Ga$_{1-x}$Al$_x$)$_4$. Our data instead indicate that the maximal THE in Eu(Ga$_{1-x}$Al$_x$)$_4$ arises from interactions of itinerant electrons with frustrated spin fluctuations in a topologically trivial magnetic state.

## II. EXPERIMENTAL METHODS

Single crystals of Eu(Ga$_{1-x}$Al$_x$)$_4$ with $x = 0.9$ and 1 were grown using a self-flux technique described in [36]. Details on the magnetization and transport measurements are described in [35]. Details for the analysis of the THE of the $x = 1$ can be found in [31]. Magnetization measurements were made using a Quantum Design Dynacool system with a vibrating sample magnetometer (VSM) option. Four-quadrant longitudinal and Hall resistivity measurements were performed using a Quantum Design Dynacool system with an electrical transport option and the data were symmetrized or antisymmetrized, respectively.

Several neutron scattering measurements were performed at Oak Ridge National Laboratory, including at the elastic diffuse scattering spectrometer CORELLI at the Spallation Neutron Source, and at the Versatile Intense Triple-Axis Spectrometer (VERITAS) and General-Purpose Small-Angle Neutron Scattering Diffractometer (GP-SANS) at the High Flux Isotope Reactor [38, 39]. Nuclear and magnetic Bragg peaks collected during measurements using the CORELLI spectrometer were integrated using the Mantid software [40]. The crystal refinement software Jana [41] was used to refine the structural and magnetic structures. Measurements at CORELLI and VERITAS used a 3.6 mg, single crystal piece cut into a rod-like shape and oriented in the ($H, 0, L$) scattering plane. The thickness of the single crystal was ~0.4 mm, which was optimized for the transmission geometry due to the strong absorption of the Eu element. A vertical field magnet was used to apply an in-plane magnetic field. Data were collected at temperatures between 1.6-300 K and magnetic fields between 0-5 T. Neutron diffraction data were fit using Rietveld refinement with the program FULLPROF [42]. Measurements at GP-SANS used a plate-like crystal, and a horizontal magnet was used to apply a magnetic field along the $c$-axis.

## III. RESULTS & DISCUSSION

### A. In-Plane Magnetic Field

First, the magnetic structures of Eu(Ga$_{1-x}$Al$_x$)$_4$ for $x = 0.9$ with an in-plane magnetic field were investigated via neutron diffraction. Figures 2(a-c) show the distinct magnetic phases, ICM, ICM+FM, and forced ferromagnetic (FFM), observed. Figure 2(a) shows the neutron diffraction pattern obtained at 2 K, 0 T. Here, four distinct ICM magnetic peaks are observed in the ($H, K, 1$) plane at $H, K = \pm$ ~0.2 r.l.u.. Since these results are similar to previous neutron scattering work on EuAl$_4$ [14], we assume the underlying magnetic structure of the $x = 0.9$ compound is similar to that of the undoped one. To accurately determine the magnetic structure of the $x = 0.9$ compound, we carried out wide-angle diffraction measurements over a large region of reciprocal space using CORELLI. Peak intensities were extracted, and the magnetic structure was refined. The refinement results indicate that the ICM peaks correspond to a screw spin helical structure, shown in Fig. 2(d), with one domain propagating along the [$H, 0, 0$] direction and the other along the [$0, K, 0$] direction. Our result is consistent with the screw spin helical structure previously identified in the parent compound, EuAl$_4$, via SANS measurements with a $c$-axis magnetic field [14]. This previous work identified four distinctive magnetic phases (I, II, III, and IV) in EuAl$_4$ where phase I exists at zero-field and forms a screw spin helix with moments aligned ferromagnetically within the ($H, 0, L$)-plane, but neighboring spins rotate within the ($H, 0, L$)-plane perpendicular to $Q$, which is directed along the [$0, K, 0$] direction [14]. As in the $x = 0.9$ Eu(Ga$_{1-x}$Al$_x$)$_4$ sample, two domains rotated by 90 degrees exist in near equal populations, giving rise to the four distinct magnetic peaks in the SANS pattern in EuAl$_4$. The screw structure can propagate along in-plane directions due to broken symmetry with structural distortion within the lattice plane [43]. X-ray diffraction experiments in EuAl$_4$ have previously revealed a structural distortion coupled with the magnetic ordering below $T_N$ [44, 45], similar to what was observed in another member of the series, EuAl$_2$Ga$_2$ [18]. In EuAl$_4$, the AFM transition is accompanied by a structural transition with a discontinuous change in lattice constants *a* and *b* [44, 45]. Orthorhombic structural distortion of 0.03-0.1% within the *ab*-plane was associated with the skyrmion phases in EuAl$_4$ [44, 45]. More recent work suggests that the crystal structure of EuAl$_4$ below CDW transition may be tetragonal noncentrosymmetric [46]. This would allow the DM interactions, a prerequisite for many skyrmion lattice hosts [8].

As a magnetic field is applied along the [0, 1, 0] direction, Eu(Ga$_{1-x}$Al$_x$)$_4$ with $x = 0.9$ undergoes a phase transition near 0.3 T corresponding to the transition from the ICM to ICM+FM magnetic states. As the magnetic field is further increased to ~1.5 T, the material undergoes a phase transition from the ICM+FM to a FFM state. These phase transitions are observed in the Hall resistivity $\rho_{yx}$ measurements shown in Fig. 3(a) and the magnetization measurements summarized in the phase diagram shown in Fig. 3(b). The Hall resistivity $\rho_{yx}$ for $H||b$ shown in Fig. 3(a) shows only an anomalous Hall contribution, in contrast to the effect measured when $H||c$ [35]. This distinction is most evident in the inset of Fig. 3(a), where $T = 2.5$ K and the extrapolation from the high magnetic field data has a nonzero $\rho_{yx}$ intercept. Fig. 3(b) presents the *H-T* phase diagram derived from d$M$/d$H$ (triangles) and d($MT$)/d$T$ (squares).

These phase transitions are consistent with those observed in the neutron diffraction measurements. With an in-plane magnetic field, the intensity of the ICM peaks along the [$H$, 0, 0] direction weakens while the ICM peaks along the [0, $K$, 0] direction strengthen, as shown in Fig. 3(c). Above ~0.4 T, only one domain persists, as demonstrated by the absence of two of the four magnetic peaks shown in Fig. 2(b). As the magnetic field strength further increases, magnetic intensity is gained at the CM Bragg peak site. This corresponds to the helical structure gaining a *b*-axis FM component along the applied magnetic field direction and results in an ICM+FM conical magnetic structure, as shown in Fig. 2(e). The intensity of the ICM peaks decreases with increased magnetic field and temperature until they disappear in the FFM and paramagnetic (PM) states, as shown in Fig. 3(c-d). In these states, the magnetic intensity is concentrated entirely at the nuclear site, as shown in the FFM state at 2 K, 2 T in Fig. 2(f).

### B. Out-of-Plane Magnetic Field

Next, the magnetic phase diagrams of Eu(Ga$_{1-x}$Al$_x$)$_4$ for $x = 1$ and 0.9 with an out-of-plane magnetic field were investigated. Figures 1(c) and (e) illustrate the *H-T* phase diagrams for $x = 1$ and 0.9, respectively. These diagrams were derived from isothermal magnetization data $M(H)$ (triangles) and temperature-dependent magnetic susceptibility $M(T)$ (squares). The contour maps are obtained from the Hall resistivity subtraction, $\rho_{yx}^T$, with $H||c$ and $j||ab$. The corresponding analyses can be found in Figs. S2 and S3. In Fig. 1(c) for $x = 1$, six distinct magnetic phases are identified, in agreement with previously reported work on EuAl$_4$ [14, 31]. Interestingly, the region exhibiting the maximal THE in EuAl$_4$ does not correspond to the phases II and III, where skyrmion spin textures, square and rhombic, were previously reported [14]. Comparing the phase diagram of $x = 1$, Fig. 1(c) to $x = 0.9$, Fig. 1(e), a dramatic change is observed. At low temperatures, the number of phases is reduced from four to three, and a maximal THE is observed in phase ii. To determine if the maximal THE corresponds to a skyrmion lattice in $x = 0.9$, we performed SANS measurements, as shown in Fig. 4 and 5. By probing the region of phase space where the THE emerges, we uncover the relationship between magnetism and the topological electronic effects present in this system.

The phase II rhombic skyrmion lattice and phase IV vortex lattice observed in EuAl4 are absent in the $x = 0.9$ sample [14]. In phases II and IV of the $x = 1$ sample, eight distinct magnetic reflections were previously observed via SANS with non-orthogonal $Q$ directions, indicating the lost four-fold symmetry of the rhombic skyrmion and vortex lattices [14]. These phases exist at low temperatures in EuAl4, while at 2 K in the $x = 0.9$ sample, we observe no additional magnetic reflections besides the four peaks that give rise to the screw helix spin texture, as shown in Fig. 4(a-c). Rather than a field-induced skyrmion lattice, a magnetic field can induce the screw helix to undergo a transition to an approximately sinusoidal fan-like structure when a magnetic field is applied perpendicular to the helix rotation axis [47]. Therefore, we propose that the zero-field screw spin helix with moments in the $[H, 0, L]$-plane transitions to a fan-like structure with non-collinear spins with a $c$-axis magnetic field. Similar transitions have been observed in other materials, including YMn6Sn6, where spin fluctuations give rise to a large THE [23-25].

A small amount of Ga-doping drastically changes the topological spin textures present in EuAl4. The Phase III square skyrmion lattice observed in EuAl4 is also present in the $x = 0.9$ sample, although it emerges most strongly in a different region of temperature-magnetic field phase space. In both EuAl4 and the $x = 0.9$ compounds, the square skyrmion lattice is experimentally observed as a distinct, new peak that emerges with the application of a magnetic field. In both compounds, the sSkL peak emerges at the $Q_1 + Q_2$ position, where $Q_1$ and $Q_2$ are the ICM Peak locations [14]. In the $x = 0.9$ sample, $Q_1$ and $Q_2$ are at (~0.15, ~0.15, 0) and (~0.15, ~-0.15, 0), respectively, and the $Q_1 + Q_2$ position of the sSkL peak is at (~0.3, 0, 0), as shown in Fig. 4(e). As shown in Fig. 1(e) for $x = 0.9$, circles denote where SANS measurements observed the field-induced sSkL peak, with the color intensity reflecting the relative intensity of the distinct sSkL peak. While the sSkL peak exists even at 2 K in EuAl4, the sSkL lattice peak is not observed at low temperatures in the $x = 0.9$ sample, as shown in Fig. 4(a-c). Rather, the sSkL lattice only emerges in a pocket of phase space at high temperatures and under a magnetic field, as shown in Fig. 1(e-f), Fig. 4(e), and Fig. 5(c-d).

The sSkL gives rise to a modest THE as shown in Fig. 1(e). Yet, a maximal THE, approximately twice as large, is observed at low temperatures where the sSkL lattice is absent. This is also the case in the parent compound EuAl4, where the maximal THE does not occur where SkLs are observed and is approximately twice as large in magnitude as the THE in the sSkL region (Fig. 1(c)). Skyrmion lattices are not a requirement of the THE. Instead, noncoplanar but topologically trivial spin textures can give rise to a THE response if they have non-zero scalar spin chirality. The net scalar spin chirality, $\chi^{sc}$, can be estimated on the tetragonal lattice by

$$\chi^{sc} = \left[\frac{1}{N}\sum_{i,\delta=\pm 1} S_i \cdot (S_{i+\delta\hat{x}} \times S_{i+\delta\hat{y}})\right]$$

where $N$ is the system size, $i$ is the position site, $S_i$ is the localized spin, and $\hat{x}, \hat{y}$ are unit vectors in the $x, y$ directions [48, 49]. The square skyrmion lattice gives rise to a net scalar spin chirality and, therefore, a topological Hall component. In contrast, the fan-like magnetic structure above the metamagnetic transition near 1 T does not give rise to a net scalar spin chirality but still yields a large topological Hall-like effect. Such a fanlike structure has contributed to a THE in other materials, such as YMn$_6$Sn$_6$ [23-25], where the magnetic structure, in combination with spin fluctuations, supports a topological Hall contribution. Several other noncollinear and collinear structures with spin fluctuations have also been shown to give rise to a THE-like response, such as Gd$_3$Ru$_4$Al$_{12}$ [50] and ErMn$_6$Sn$_6$ [51]. While the scalar spin chirality in these systems is equal to zero, spin fluctuations from the geometrical frustrated antiferromagnets can also contribute to the AHE by scattering the itinerant electrons even in the spin disordered state [52-54]. Unlike YMn$_6$Sn$_6$, the largest THE in Eu(Ga$_{1-x}$Al$_x$)$_4$ with $x = 0.9$ is found at the lowest temperatures, indicating another mechanism must be responsible for the effect. Recently, a metamagnetic Hall effect has been presented in ErGa$_2$ with a magnitude of ~0.06 μΩ cm [26]. The effect size in Eu(Ga$_{1-x}$Al$_x$)$_4$ with $x = 0.9$ is comparable in magnitude. Therefore, we conclude that the maximal THE-like response in Eu(Ga$_{1-x}$Al$_x$)$_4$ may similarly not have topological origins. Instead, our data suggest that the maximum THE in Eu(Ga$_{1-x}$Al$_x$)$_4$ is likely associated with interactions of itinerant electrons with frustrated spin fluctuations arising from a noncollinear magnetic ordered state (Fig. 4).

Since the maximal THE and sSkL lattice do not overlap in phase space, we consider different methods for the identification of topological spin textures. A possible identification measurement is the MCE since the interplay of magnetic anisotropy and electric polarization induced by the spiral planes and skyrmion lattice leads to large magnetocaloric effects [55]. The MCE has been shown to be an effective method for determining the phase diagrams of skyrmion materials, e.g., in the cubic phase of FeGe [37]. These measurements allow for quantifying the latent heat associated with magnetic-field-driven phase transitions.

From the Maxwell relation, $\left(\frac{dS}{dH}\right)_T = \left(\frac{dM}{dT}\right)_H$ where $S$ is the total entropy, the MCE is determined as the change in magnitude of the isothermal entropy with changing magnetization, $\Delta S_M(T, H)$, given by

$$\Delta S_M(T, H) = \int_0^H \left(\frac{dM}{dT}\right)_{H'} dH'$$

This formulation allows for the direct determination of the isothermal entropy obtained from bulk DC magnetization measurements. Figure 6(a-b) shows the magnetization $M(T)$ measured under different applied magnetic fields, where the field was measured at intervals of 50 Oe. This high density of data points enables detailed observation of the entropy changes, as seen in Fig. 1(f) for $x = 0.9$. Following the $M(T)$ measurement, $dM/dT$ can be calculated, providing a map of $\Delta S_M$. Fig. 6(c-d) presents the $H$-$T$ phase diagrams for $dM/dT$ for $x = 1$ and 0.9, respectively. Since $dM/dT = dS/dH$, blue regions correspond to a negative change in entropy and indicate polarizing spins. In contrast, white regions indicate near-zero entropy change and dark red indicates regions of enhanced entropy, which occur near a phase transition.

The MCE results are summarized in the phase diagrams in Fig. 1(d) and (f) for $x = 1$ and 0.9, respectively. The MCE identifies phase transitions, indicated by changes in first-order phase boundaries, that can correspond to topologically distinct spin states. The MCE analysis for $x = 0.9$ shows the largest entropy change occurring between phases ii and iii, as depicted in Fig. 1(f), coinciding with the presence of the sSkL peak observed via SANS measurements. Similarly, in the $x = 1$ sample, the largest entropy change corresponds to the topologically distinct skyrmion phases II and III, as depicted in Fig 1(d). The dashed lines represent first-order phase transitions.

The observation of a maximum in THE is not necessarily associated with the location of a skyrmion lattice. Indeed, in both the $x = 1$ and 0.9 samples, the maximal THE does not correspond to the skyrmion lattice phases. Rather, a strong MCE identifies the boundary between the topologically trivial and nontrivial spin structures [56]. Therefore, we conclude a large MCE rather than a maximal THE response is a strong indicator of the topologically nontrivial magnetic phases in the Eu(Ga$_{1-x}$Al$_x$)$_4$ system. The MCE provides an alternative method to map phase diagrams of skyrmion hosts and can be considered in other systems to identify skyrmion lattice boundaries, especially those hosting large topological Hall-like responses, potentially arising from topologically trivial spin textures.

## IV. CONCLUSION

In summary, we have used magnetotransport, magnetization, and neutron scattering to map out the *H-T* phase diagram of Eu(Ga$_{1-x}$Al$_x$)$_4$ ($x = 0.9$). The neutron diffraction results identify several distinct magnetic phases that arise when the field is applied either in-the-plane or along the *c*-axis. This includes a sSkL, identical to that previously reported in EuAl$_4$. At zero-field, a screw helical spin structure exists with two domains propagating along the [*H*, 0, 0] and [0, *K*, 0] directions. Applying an in-plane field populates one domain at the expense of the other and cants the magnetic structure in the field direction resulting in a conical ICM+FM phase before becoming fully polarized in the FFM phase. A *c*-axis magnetic field gives rise to large magnetic and electronic responses, including a topological Hall response. Our neutron diffraction results show that the maximal topological Hall-like response in Eu(Ga$_{1-x}$Al$_x$)$_4$ ($x = 1, 0.9$) corresponds to a topologically trivial, fan-like magnetic state where the large electronic response must be attributed to another mechanism such as interactions of itinerant electrons with frustrated spin fluctuations. Additionally, a sSkL exists in Eu(Ga$_{1-x}$Al$_x$)$_4$ ($x = 0.9$) but occupies a distinct phase space area when compared to the $x = 1$ sample. In both the $x = 0.9$ and 1 compounds, there is a discrepancy between the sSkL phase and the maximal THE. We conclude that while a small THE arises due to the skyrmion lattice, the sSkL phase can be identified more accurately by the maximal MCE, and the maximal THE must arise from itinerant electron-spin fluctuations interactions in a topologically trivial magnetic state.

**ACKNOWLEDGMENTS**
The neutron scattering and materials synthesis work is supported by the US DOE, Basic Energy Sciences (BES), under grant no. DE-SC0012311 and by the Robert A. Welch Foundation under grant no.C-1839, respectively (P.D.). K.J.N. is in part supported by the U.S. Department of Energy, Office of Science, Office of Workforce Development for Teachers and Scientists, Office of Science Graduate Student Research (SCGSR) program. The SCGSR program is administered by the Oak Ridge Institute for Science and Education for the DOE under contract number DE-SC0014664. KJA and EM have been supported by the Robert A. Welch Foundation under grant no.C-2114 and JM was partially supported by the NSF Grant NSF DMR 1903741. A major portion of this research used resources at the High Flux Isotope Reactor and Spallation Neutron Source, a DOE Office of Science User Facility operated by the Oak Ridge National Laboratory.

Appendix:

## I. MAGNETIZATION AND TRANSPORT MEASUREMENTS

### Figure S1: Eu(Ga$_{1-x}$Al$_x$)$_4$ Magnetization with $H||c$

Fig. S1(a-b) shows the measured magnetization, $M(H)$, curves at different temperatures. The corresponding $H$-$T$ phase diagrams are determined from the isothermal magnetization measurements are shown in Fig S1(c-d). The magnetic phase boundaries for Fig. 1(c) and 1(e) are obtained from the derivative in d$M$/d$H$, consistent with previous studies [35]. The magnetic moment saturates at 7 $\mu_B$/Eu in the spin-polarized (SP) phase above 2 T and 2 K for both $x = 0.9$ and 1.

### Figure S2: Eu(Ga$_{1-x}$Al$_x$)$_4$ with $x = 0.9$ topological Hall effect (THE) analysis

The THE analysis for $x = 0.9$ is consistent with that reported in [35]. Evaluating the THE in $x = 0.9$ is challenging due to the nonlinear field dependence above $H_c$, where the $M$ is saturated. This composition requires a multiband description of the normal Hall effect to fully account for the normal, anomalous (AHE), and THE behavior. For this case, we note that in the high field region, $H > 8$ T, the hall resistivity magnetic field dependence becomes that of a two-band model. This enables a two-band model subtraction to extract the AHE and THE dependence following $\Delta\rho_{yx} = \rho_{yx} - \rho_{two-band} - \rho_{yx}^{AHE} - \rho_{yx}^{THE}$.

### Figure S3: Eu(Ga$_{1-x}$Al$_x$)$_4$ with $x = 1$ topological Hall effect (THE) analysis

Similar to the $x = 0.9$ case, EuAl$_4$ presents a multiband effect in addition to quantum oscillations for $H > 8$ T, as shown in the inset of Fig. S3(b). In this case, we follow the analysis detailed in [31]. This includes subtracting a polynomial background, as shown in Fig. S3(c). The obtained $\Delta\rho_{yx}$ (Fig. S3(d)) contains part of the AHE that could be trivial (conventional AHE) or nontrivial (THE) in nature.

### Figure S4: Eu(Ga$_{1-x}$Al$_x$)$_4$ with $x = 0.9$ magnetization with $H||ab$

Magnetization measurements were made to define the boundaries in the phase diagram with an in-plane magnetic field, as shown in Fig. 3(b) in the main text. The details are included here. Fig. S4 shows the $M(H)$ curves measured at different temperatures. The magnetic phase boundary for Fig. 3(c-f) in the main text is obtained from the derivative in d$M$/d$H$, as shown in Fig. S4(b).

### Figure S5: Eu(Ga$_{1-x}$Al$_x$)$_4$ with $x = 0.9$ susceptibility with $H||ab$

Fig. S5(a) shows the $M(T)$ curves measured at different magnetic fields. The magnetic phase boundary for Fig. 3b in the main text is obtained from the derivative in d$MT$/d$T$, as shown in Fig. S5(b).

## II. NEUTRON DIFFRACTION MEASUREMENTS AND DATA ANALYSIS

**Figure S6: Photograph of the Eu(Ga$_{1-x}$Al$_x$)$_4$ with $x = 0.9$ sample measured during CORELLI measurements**

Due to the large absorption of Eu, a single crystalline piece of Eu(Ga$_{1-x}$Al$_x$)$_4$ with $x = 0.9$ sample was cut into a thin rod, as shown in Fig. S6. Therefore as the sample was rotated 360 degrees to collect a reciprocal space map, a uniform cross-section was irradiated with neutrons.

**Figure S7: Refinement results for Eu(Ga$_{1-x}$Al$_x$)$_4$ with $x = 0.9$**

Nuclear and magnetic Bragg peaks collected during measurements using the CORELLI [38] spectrometer were integrated using the Mantid software [40]. The crystal refinement software Jana was used to refine the structural and magnetic structures. The observed intensities compared to the expected intensities for the refined structures are shown in Fig. S7.

FIGURES:

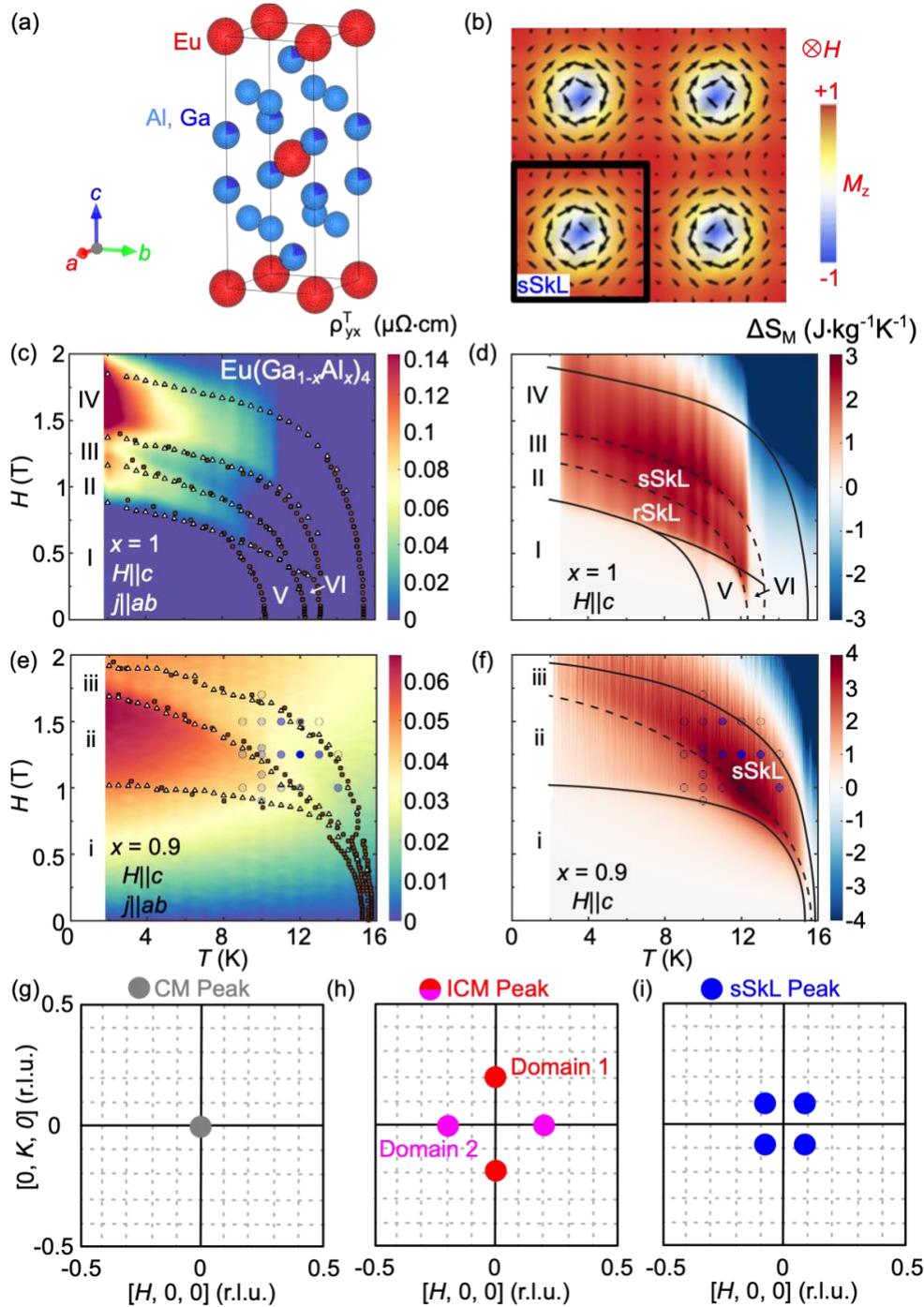

Fig. 1. (a) Crystal structure of Eu(Ga$_{1-x}$Al$_x$)$_4$. (b) Schematic of the square skyrmion lattice (sSkL) [14]. (c) Eu(Ga$_{1-x}$Al$_x$)$_4$ magnetic field-temperature ($H$-$T$) phase diagram with contour map corresponding to the contribution to the anomalous and topological Hall resistivity $\rho_{yx}^T$ for $x = 1$. Beige triangles were determined by maxima in d$M$/d$H$ and brown squares were determined from d($MT$)/d$T$ measurements. (d) Phase diagram with contour maps of $\Delta S_M(T, H)$ for $x = 1$. The phase regions separated by features in d$S$/d$H$ are visualized in terms of entropy. The lines indicate the phase boundaries. The dotted lines show regions where first-order phase transitions

occur and correspond to the rhombic and square skyrmion lattices. (e) Eu(Ga$_{1-x}$Al$_x$)$_4$ magnetic field-temperature (*H-T*) phase diagram with contour map corresponding to the contribution to the anomalous and topological Hall resistivity $\rho_{yx}^T$ for *x* = 0.9. Blue circles indicate regions where the sSkL peak in small-angle neutron scattering (SANS) measurements was observed with the relative intensity of the peak indicated by the shading of the circle. (f) Phase diagram with contour maps of $\Delta S_M(T, H)$ for *x* = 0.9. (g-i) Schematic of the [*H, K*, 0] plane reciprocal space maps showing (g) commensurate (CM) peaks, (h) incommensurate (ICM) peaks, and (i) square skyrmion lattice (sSkL) peaks with grey, red/pink, and blue dots, respectively.

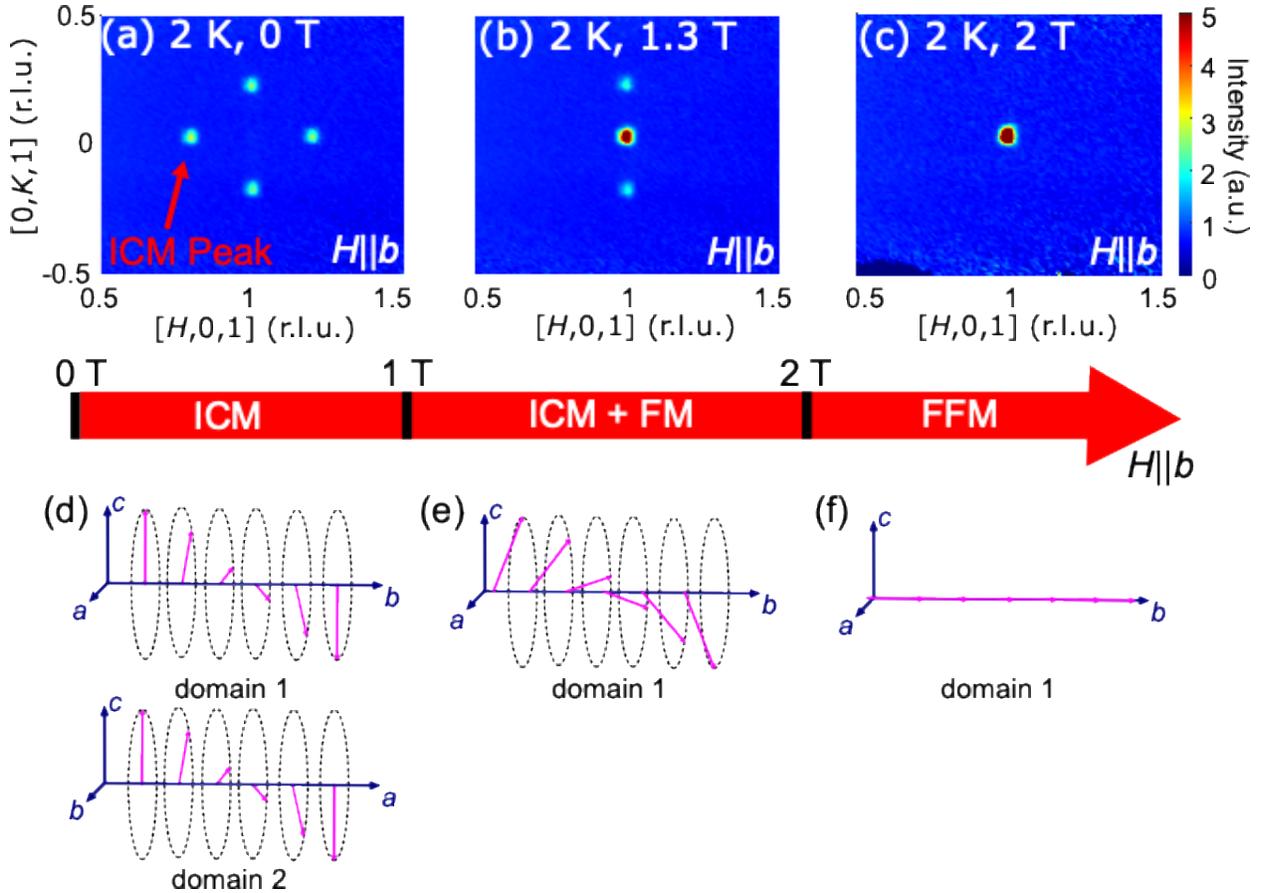

Fig. 2. (a-c) Neutron diffraction patterns about the (1,0,1) peak as a function of the magnetic field. (a) 2 K, 0 T has two magnetic domains corresponding to the peaks along *H* and *K*. (b) At 2 K, 1.3 T, only one magnetic domain with peaks along *K* persists. (c) At 2 K, 2 T, the ICM peaks disappear. (d-f) Magnetic structure of Eu(Ga$_{1-x}$Al$_x$)$_4$ with *x* = 0.9 as a function of an in-plane magnetic field. (d) At 0 T, the magnetic structure hosts two domains, each comprising a two-dimensional ICM helical structure. Domain one propagates along the *b*-axis, while domain two propagates along the *a*-axis. (e) At 1.3 T, only domain 1 persists, and the ICM magnetic structure gains a net FM moment along the field direction, *b*, resulting in a conical magnetic structure (ICM+FM). (f) At 2 T, the magnetic structure is a forced ferromagnet (FFM) with magnetic moments directed along the *b*-axis.

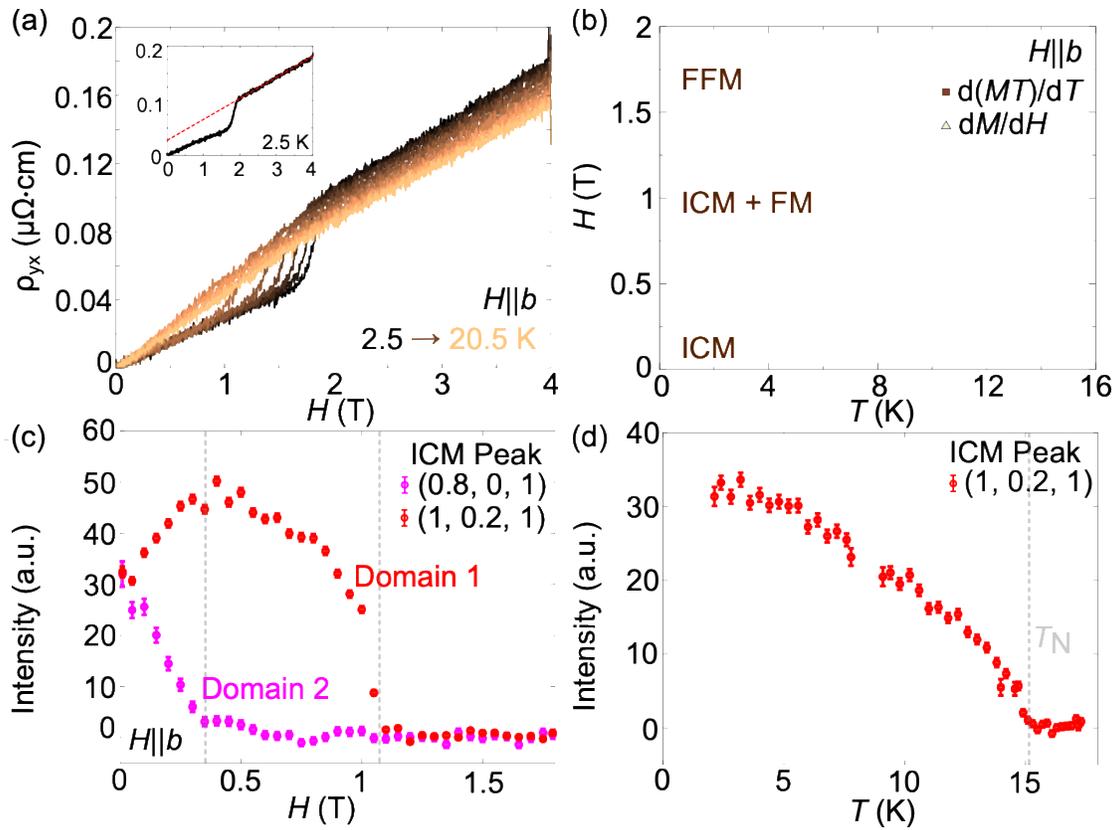

Fig. 3. (a) Hall resistivity as a function of magnetic field and temperature for $H\|b$. The inset shows the Hall resistivity curve with the high magnetic field fitting (red line) indicating the AHE behavior at 2.5 K. (b) Phase diagram obtained from maximum points in d$M$/d$H$ and d$MT$/d$T$. (c) In-plane field dependence of the incommensurate (ICM) peaks along $H$, (0.8, 0, 1) and $K$, (1, 0.2, 1) at 10 K. (d) Temperature dependence of the ICM peak along $K$, (1, 0.2, 1) at 0 T.

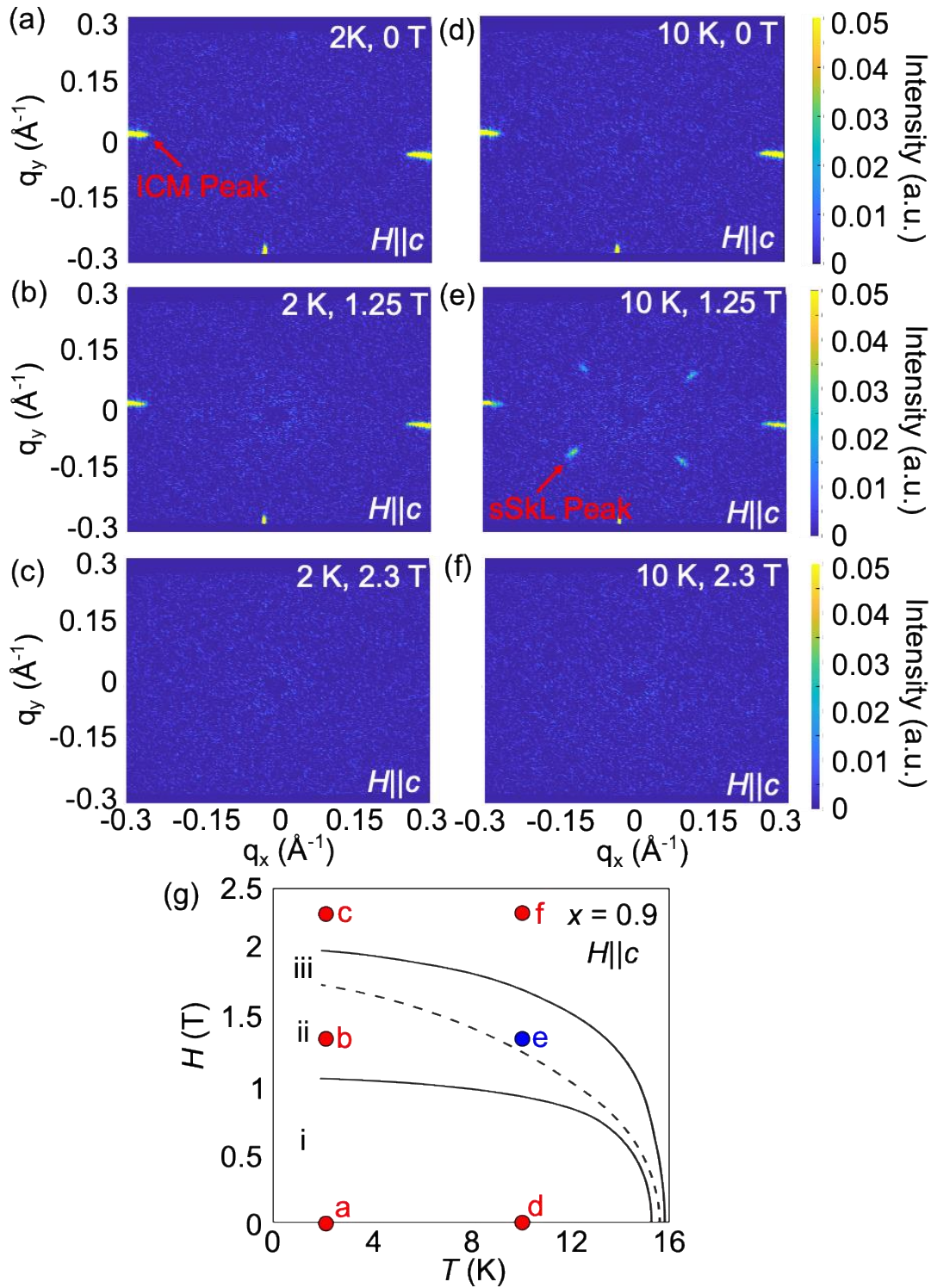

Fig. 4. Small-angle neutron scattering patterns as a function of temperature and c-axis magnetic field. (a-c) SANS patterns at 2 K with (a) 0 T, (b) 1.25 T, and (c) 2.3 T. (d-f) SANS patterns at 10 K with (d) 0 T, (e) 1.25 T, and (f) 2.3 T. The ICM peaks are also present in the neutron diffraction patterns. The square skyrmion lattice (sSkL) peak is observed at 10 K, 1.25 T. (g) Schematic of the phase diagram where circles indicate where in phase space the measurements shown in (a-f) where taken.

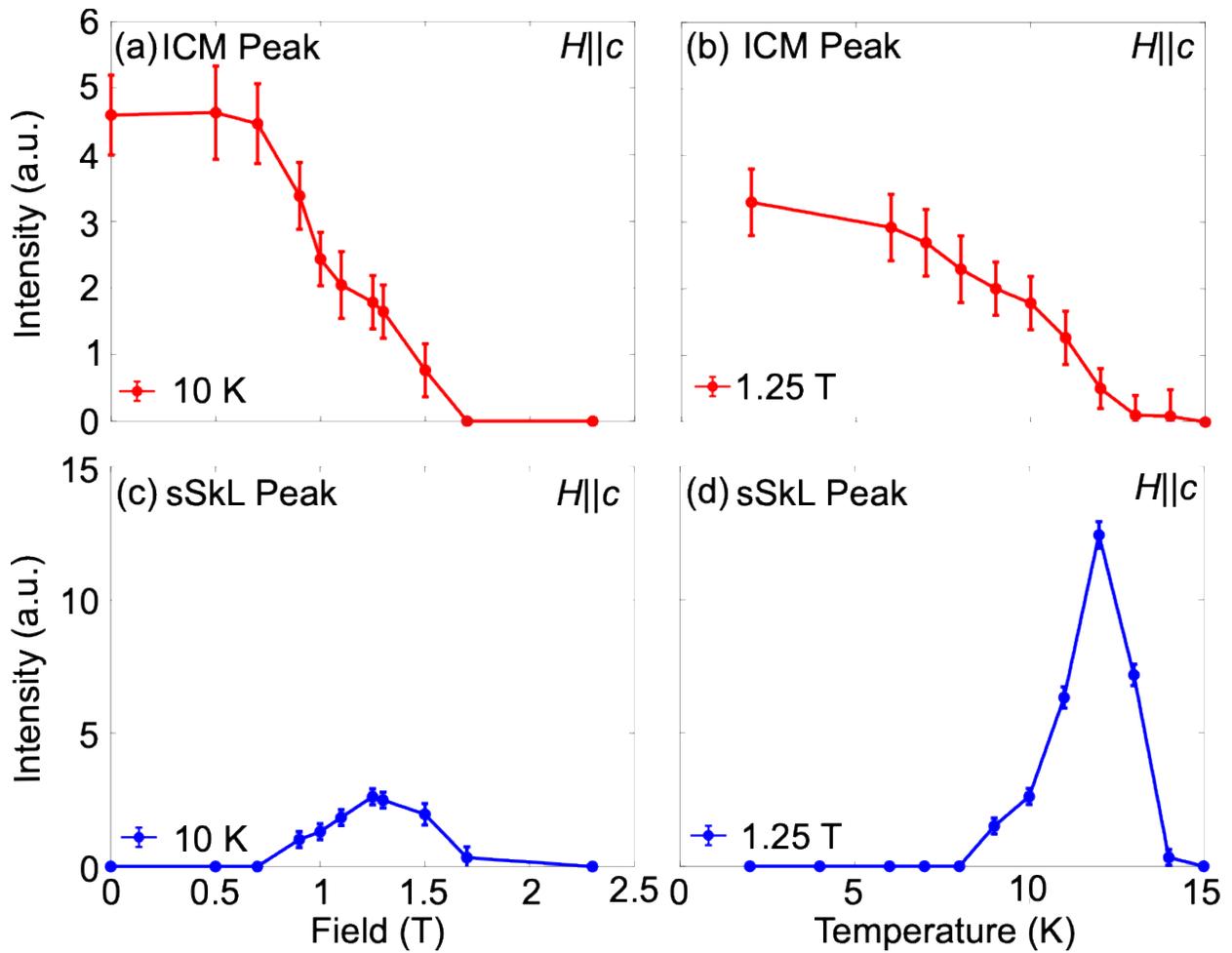

Fig. 5. (a-b) SANS integrated intensity of the ICM Peak reflection as a function of (a) *c*-axis magnetic field at 10 K and (b) temperature dependence at 1.25 T along the *c*-axis. (c-d) SANS integrated intensity of the sSkL Peak reflection as a function of (c) *c*-axis magnetic field at 10 K and (d) temperature dependence at 1.25 T along the *c*-axis.

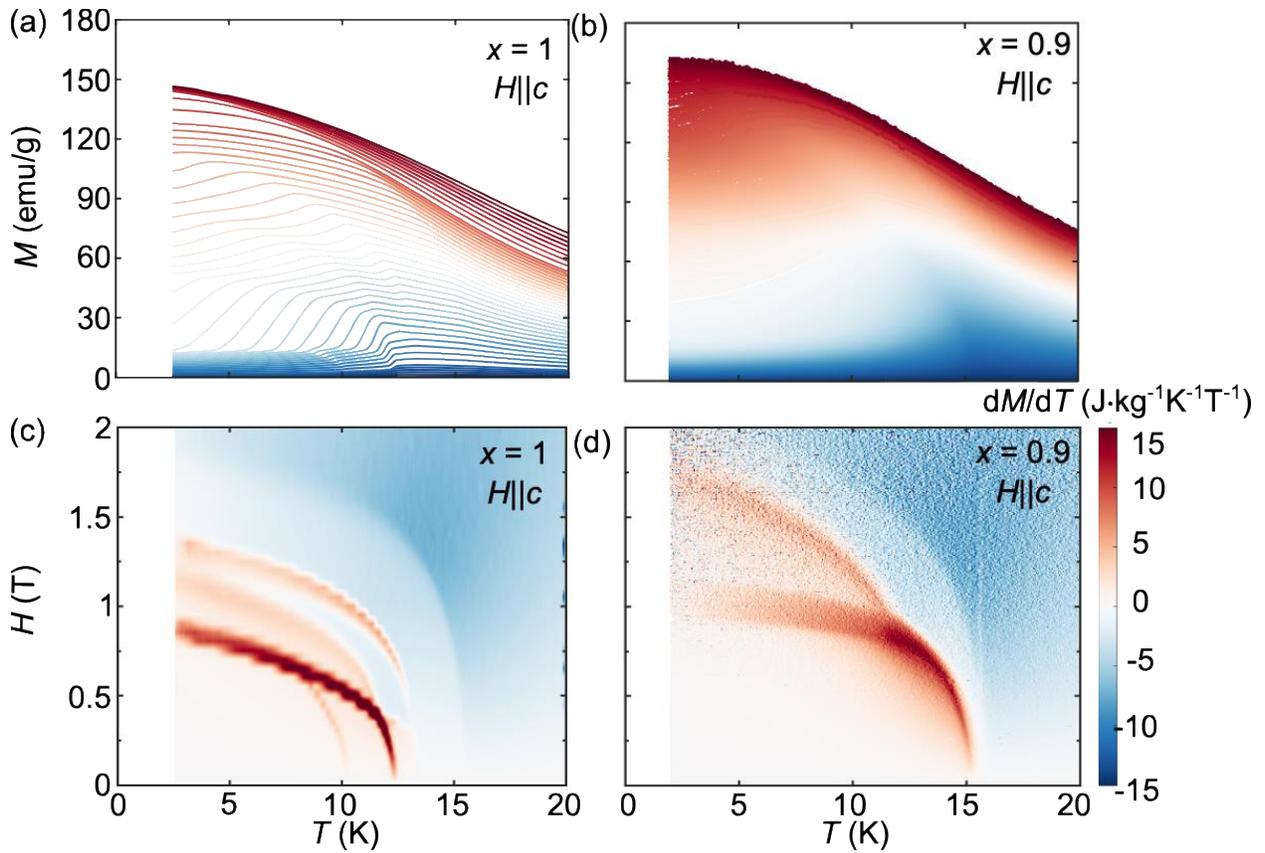

Fig. 6. Eu(Ga$_{1-x}$Al$_x$)$_4$ Magneto-caloric effect (MCE) analysis. (a-b) Moment vs temperature with magnetic field $H\|c$ for (a) $x = 1$ and (b) 0.9. The magnetic field was measured from 0.005-2 T with measurements every (b) 50 Oe to obtain a high density of data points for the MCE analysis. (c-d) Magnetic field-temperature ($H$-$T$) phase diagrams for (a) $x = 1$ and (b) 0.9. The contour maps correspond to d$M$/d$T$.

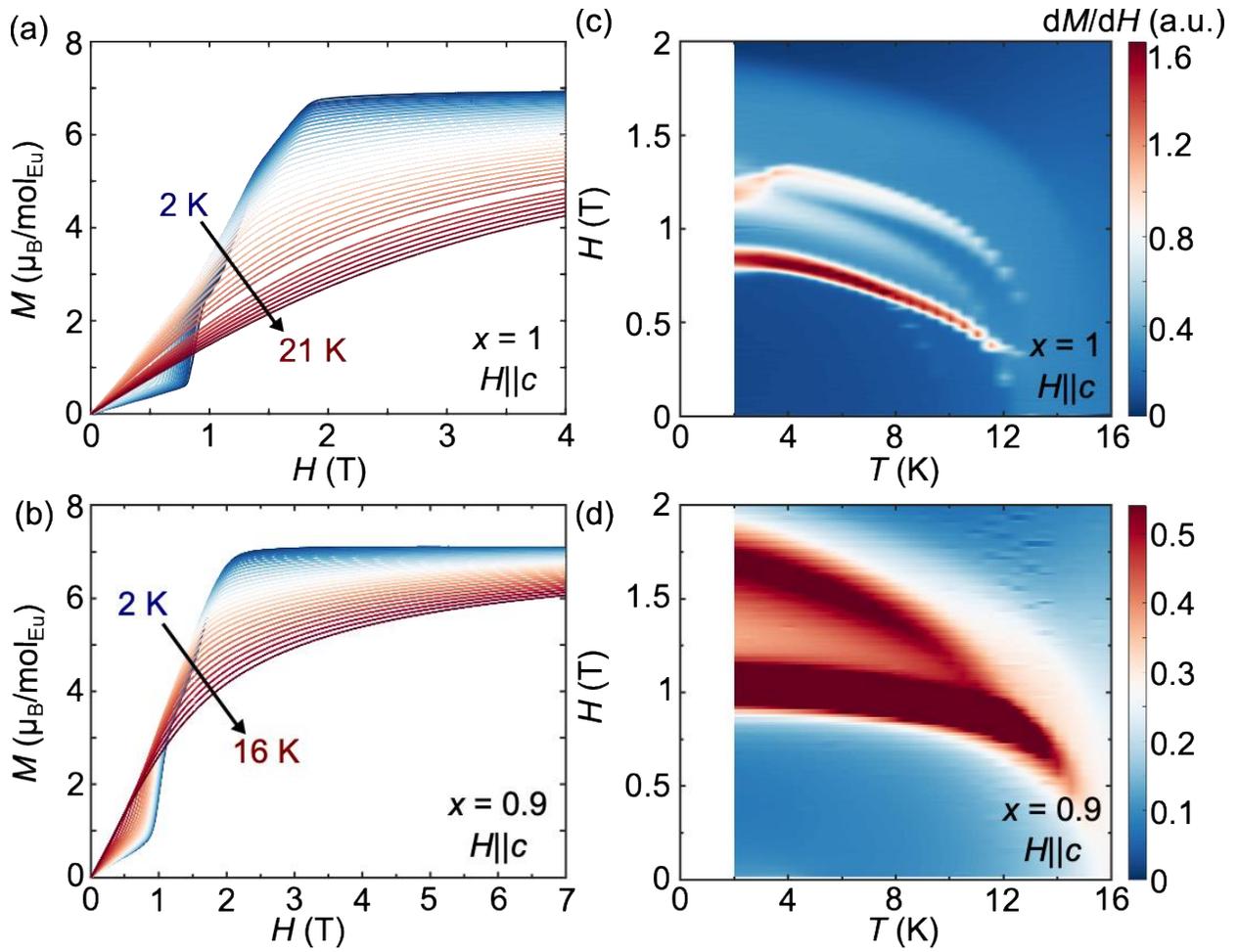

Fig. S1. Eu(Ga$_{1-x}$Al$_x$)$_4$ Magnetization with $H||c$. (a-b) Moment vs magnetic field along the *c*-axis for (a) $x = 1$ and (b) 0.9. (b) Magnetic field-temperature ($H$-$T$) phase diagrams for (c) $x = 1$ and (d) 0.9 with the contour indicating the magnitude of d$M$/d$H$.

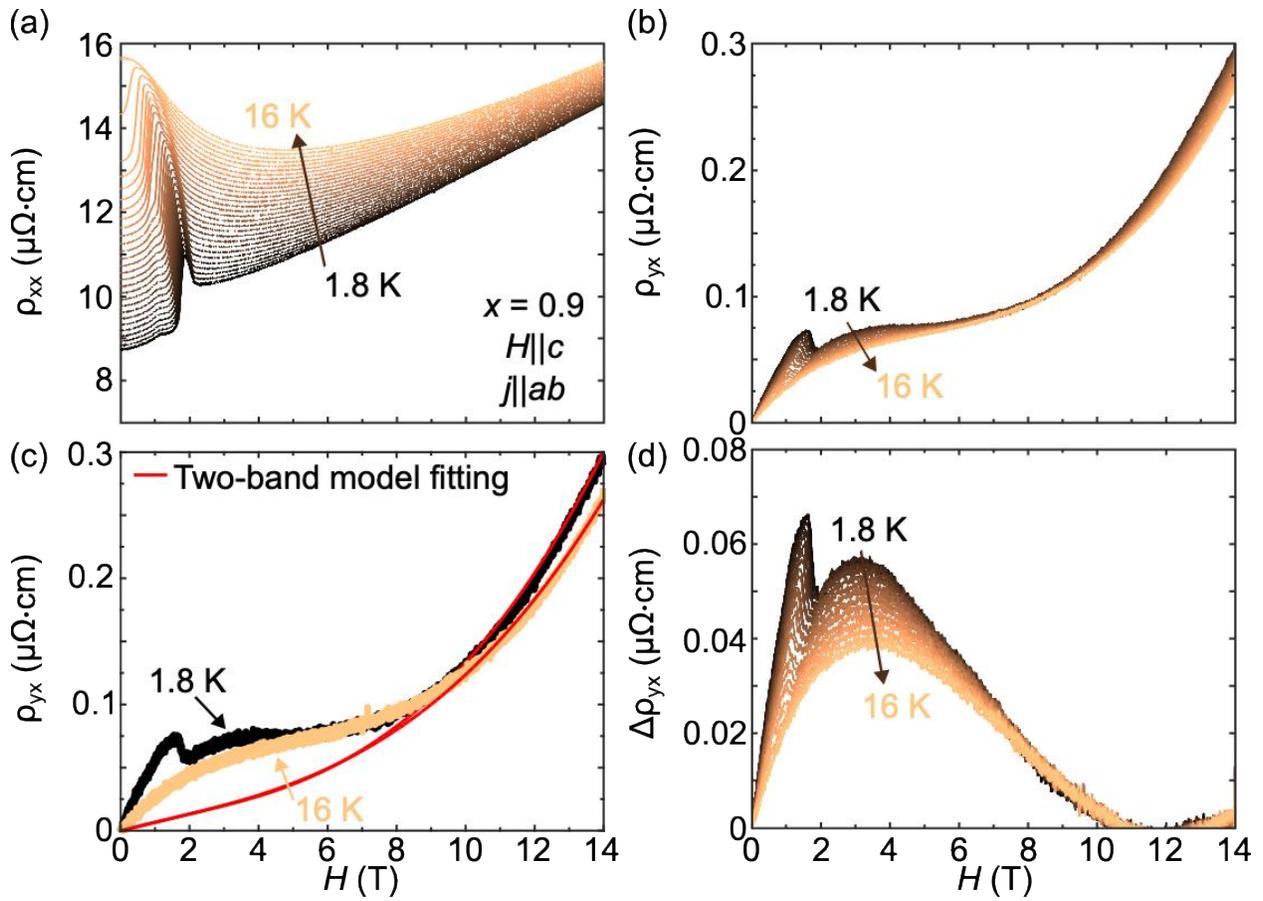

Fig. S2. Eu(Ga$_{1-x}$Al$_x$)$_4$ with $x = 0.9$ THE analysis. (a) Resistivity vs magnetic field with $H\|c$ and $j\|ab$ measured up to 14 T. (b) Hall resistivity measured up to 14 T. (c) Two-band model fit from high field for $T = 1.8$ K and above $T_N$ at $T = 16$ K. (d) THE subtraction from two-band model fitting.

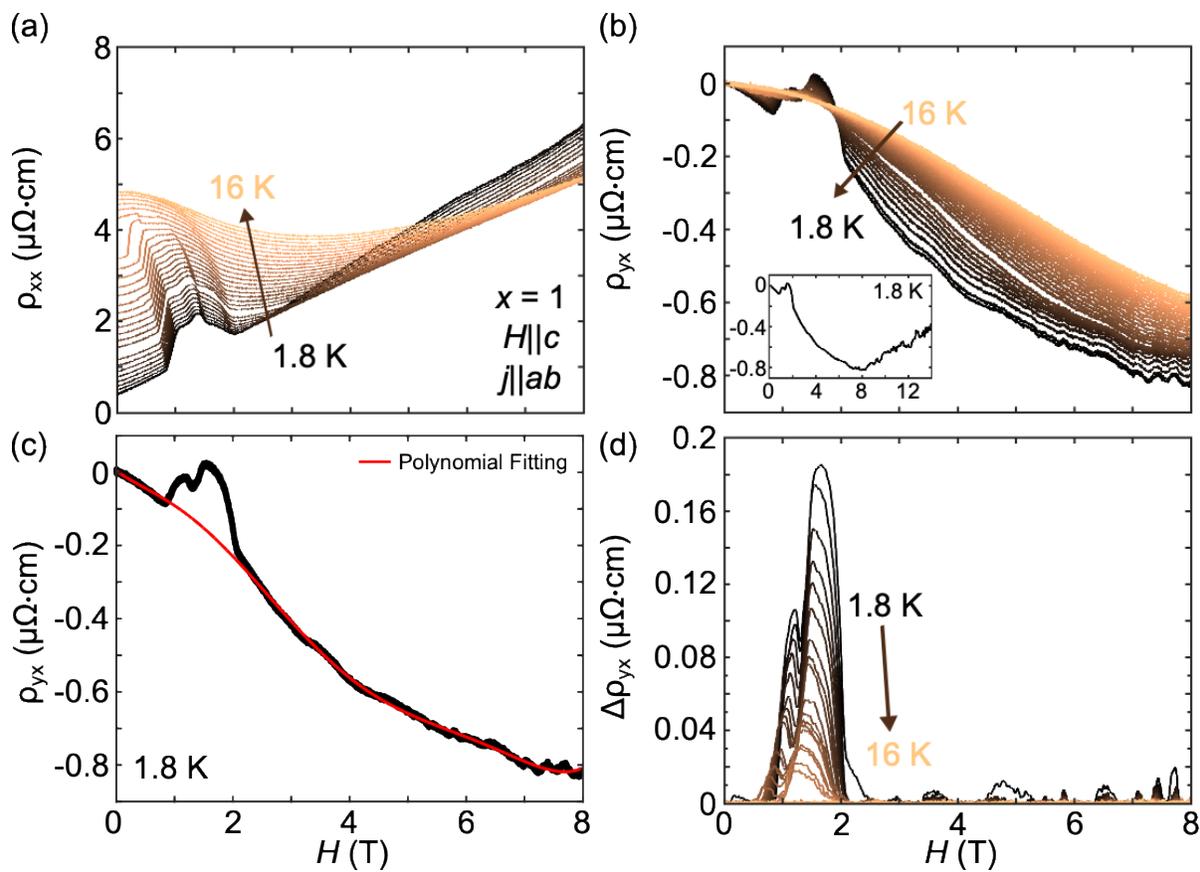

Fig. S3. Eu(Ga$_{1-x}$Al$_x$)$_4$ with $x = 1$ THE analysis. (a) Resistivity vs magnetic field with $H \| c$ and $j \| ab$ measured up to 14 T. (b) Hall resistivity measured up to 14 T. (c) Polynomial subtraction for $T = 1.8$ K up to 8 T. Accounting for multiband effect makes the subtraction more difficult in the parent x = 1 compound. (d) THE subtraction from polynomial fitting.

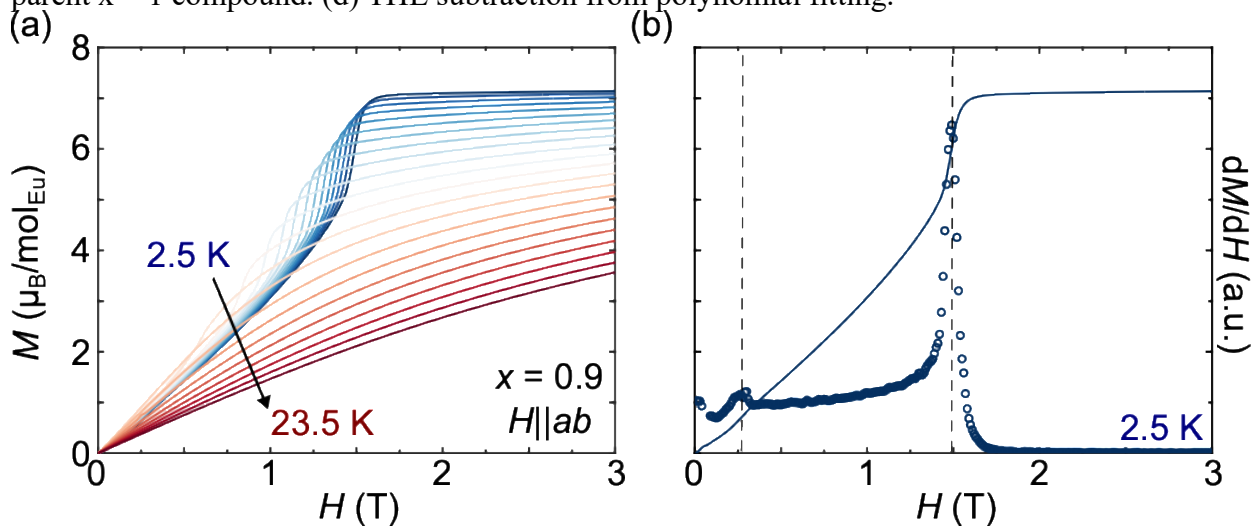

Fig. S4. Eu(Ga$_{1-x}$Al$_x$)$_4$ with $x = 0.9$ magnetization with $H \| ab$. (a) Magnetization vs field for $x = 0.9$ and $H \| ab$. (b) Determination of phase boundary points, indicated by dashed vertical lines, from the maximum in d$M$/d$H$.

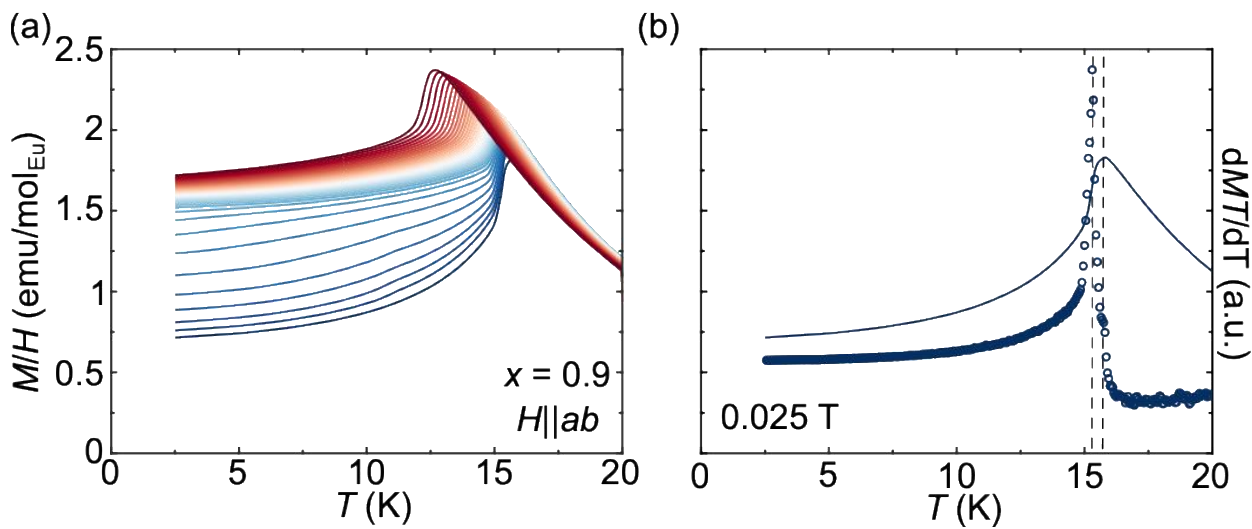

Fig. S5. Eu(Ga$_{1-x}$Al$_x$)$_4$ with $x = 0.9$ susceptibility with $H\|ab$. (a) Magnetization vs temperature for $x = 0.9$ and $H\|ab$. (b) Determination of phase boundary points, indicated by dashed vertical lines, from the maximum in d$MT$/d$T$.

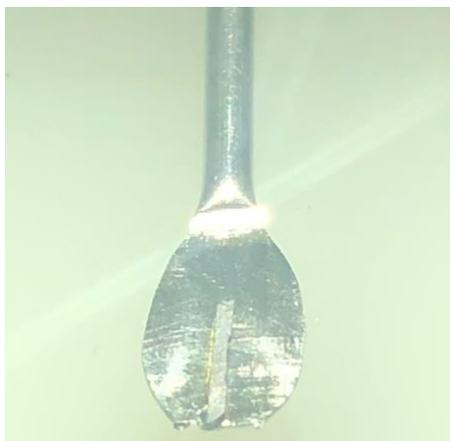

Fig. S6. Photograph of the Eu(Ga$_{1-x}$Al$_x$)$_4$ with $x = 0.9$ sample measured during CORELLI measurements. Due to the large absorption of Eu, the sample was cut into a thin rod to provide a constant sample width when the sample was rotated by 360 degrees.

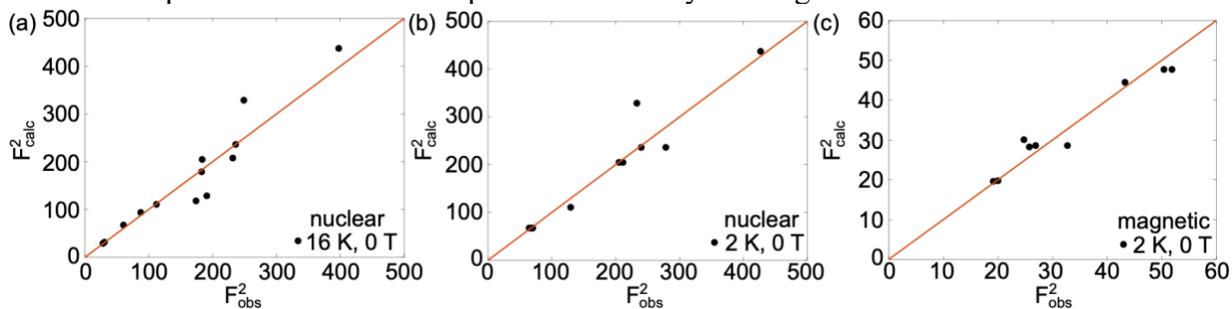

Fig. S7. Refinement results for Eu(Ga$_{1-x}$Al$_x$)$_4$ with $x = 0.9$. (a-c) Observed and calculated refinement results for (a) the nuclear structure at 16 K, 0 T, (b) the nuclear structure at 2 K, 0 T, and (c) the magnetic structure at 2 K, 0 T.

(scattering instruments at Oak Ridge National LaboratoryThis article will form part of a virtual special issue on advanced neutron scattering instrumentation, marking the 50th anniversary of the journal." *Journal of Applied Crystallography* **51**, 242-248 (2018).)